\documentclass[doublecol]{epl2}
\usepackage{amsmath,amssymb}

\newcommand{\mean}[1]{\langle #1 \rangle}
\newcommand{\pd}[2]{\frac{\partial #1}{\partial #2}}
\newcommand{\Int}[1]{\int\mathrm{d}#1\;}
\newcommand{\IInt}[3]{\int_{#2}^{#3}\mathrm{d}#1\;}
\newcommand{\unit}[1]{\;\mathrm{#1}}

\newcommand{\dd}{\mathrm{d}}
\newcommand{\trap}{\nu_\mathrm{T}}
\newcommand{\js}{j_\mathrm{s}}
\newcommand{\ps}{p_\mathrm{s}}
\newcommand{\vloc}{v_\mathrm{s}}
\newcommand{\st}{\Delta s_\mathrm{tot}}
\newcommand{\sm}{\Delta s_\mathrm{m}}
\newcommand{\fc}{f_\mathrm{c}}
\newcommand{\kB}{k_\mathrm{B}}
\newcommand{\kT}{k_\mathrm{B}T}
\newcommand{\rk}{r_\mathrm{K}}
\newcommand{\funit}{\kT/\mu\mathrm{m}}
\newcommand{\al}{\alpha}

\begin{document}

\title{Distribution of Entropy Production for a Colloidal Particle in a
  Nonequilibrium Steady State}

\author{T.~Speck\inst{1} \and V.~Blickle\inst{2} \and C.~Bechinger\inst{2}
  \and U.~Seifert\inst{1}}

\institute{
  \inst{1} II. Institut f\"ur Theoretische Physik, Universit\"at Stuttgart,
  Pfaffenwaldring 57, 70550 Stuttgart, Germany \\
  \inst{2} 2. Physikalisches Institut, Universit\"at Stuttgart,
  Pfaffenwaldring 57, 70550 Stuttgart, Germany
}

\abstract{For a colloidal particle driven by a constant force across a
  periodic potential, we investigate the distribution of entropy production
  both experimentally and theoretically. For short trajectories, the
  fluctuation theorem holds experimentally. The mean entropy production rate
  shows two regimes as a function of the applied force. Theoretically, both
  mean and variance of the pronounced non-Gaussian distribution can be
  obtained from a differential equation in good agreement with the
  experimental data.}


\date{\today}

\maketitle


Nonequilibrium steady states constitute arguably the simplest class of
nonequilibrium systems. They are characterized by a stationary distribution
but differ crucially from equilibrium states since detailed balance is broken.
As a consequence, entropy is produced at an on average positive rate.
Fluctuations of the entropy production towards negative values do occur but
they are severely constrained by the fluctuation theorem. This universal
relation was first observed in the simulations of a sheared
fluid~\cite{evan93} and later proven both for chaotic dynamic
systems~\cite{gall95} and for stochastic dynamics~\cite{kurc98,lebo99}. In
principle, the fluctuation theorem is an asymptotic statement in the long time
limit. If, however, entropy is assigned to the driven system as well and not
only to the coupled heat bath, the fluctuation theorem holds strictly for
finite times~\cite{seif05a}. Closely related to the fluctuation theorem are
the Jarzynski~\cite{jarz97} and Crooks~\cite{croo99} nonequilibrium work
relations, which proved to be useful in the determination of equilibrium free
energy differences in single molecule experiments~\cite{coll05,rito06}.

The fluctuation theorem constrains the probability of negative entropy
production. It does, however, not predict the distribution for positive
production which is, of course, a nonuniversal function. For a better
understanding of nonequilibrium steady states~\cite{zia06}, detailed studies
of the entropy production in specific systems are important.  Entropy
production has been studied both experimentally and theoretically for a
variety of systems including turbulent flows~\cite{cili04}, granular
systems~\cite{feit04}, liquid crystals~\cite{gold01}, and the ideal
gas~\cite{cleu06}, mostly addressing the entropy production in the medium
only. Entropy production including that of the system has been experimentally
measured in an athermal two-level system~\cite{tiet06} for which later
numerical calculations of its probability distribution have been
performed~\cite{impa07}. In the latter system, medium entropy is somewhat
artifically defined and should not be associated with dissipated heat.

Colloidal particles driven by time-dependent laser traps have developed into
an ideal system for quantitatively studying these new concepts in
nonequilibrium statistical mechanics for essentially two reasons. First,
individual trajectories can be traced and recorded in real space in comparison
to ensemble averages typically obtained in scattering experiments. Second,
even though the particle in a distinct nonequilibrium steady state can be
driven beyond linear response, the surrounding fluid still faithfully behaves
like an equilibrium thermal bath. In this letter, we exploit these features to
analyze entropy production in a nonequilibrium steady state consisting of a
single particle driven by a constant force across a periodic
potential~\cite{risken}. In contrast to previous experiments on colloids in
time-dependent harmonic potentials~\cite{wang02,trep04} such a periodic
potential in general leads to non-Gaussian distributions for quantities like
applied work, dissipated heat and generated entropy~\cite{mari06}. In fact, we
will show that even the mean entropy production rate shows an pronounced
crossover as a function of the applied force. An additional advantage of such
a colloidal system compared to driven bulk systems is that once the potential
is known, the experimental data can be compared to independent numerical
calculations.


\begin{figure}[t]
  \onefigure[width=.8\linewidth]{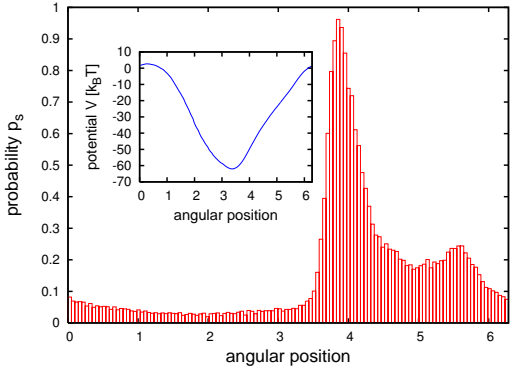}
  \caption{Stationary probability distribution $\ps(x)$ for a cosine input
    signal to the EOM with driving force $f=14.2\funit$. The inset shows the
    reconstructed actual potential $V(x)$ with $V_0=65.6\kT$.}
  \label{fig:dist}
\end{figure}

We study a single colloidal particle in a toroidal geometry driven into a
nonequilibrium steady state. The overdamped motion of the colloidal particle
is governed by the Langevin equation
\begin{equation}
  \label{eq:lang}
  \dot x(t) = \mu_0\left(-\pd{V}{x} + f\right) + \zeta(t) 
  \equiv \mu_0F(x) + \zeta(t)
\end{equation}
with a periodic potential $V(x+L)=V(x)$ and periodicity $L$. The thermal noise
$\zeta$ describes the coupling of the particle to the surrounding fluid
modeling the heat bath with temperature $T$. The noise has zero mean and
correlations $\mean{\zeta(t)\zeta(\tau)}=2D_0\delta(t-\tau)$. The correlation
strength $D_0$ of the heat bath is connected with the bare mobility of the
particle $\mu_0$ by the Einstein relation $D_0=\kT\mu_0$. The crucial
assumption is that we require the heat bath to be and to stay in equilibrium
at the constant temperature $T$. We can then identify the dissipated heat $q$
as the change of entropy
\begin{equation}
  \label{eq:sm}
  \sm[x(\tau)] = \int\frac{\dd q}{T} =
  \frac{1}{T}\IInt{\tau}{0}{t}\dot x(\tau)F(x(\tau))
\end{equation}
{\em in the heat bath or medium} along a single trajectory $x(\tau)$ of length
$t$~\cite{seif05a,blic06}. Beside this entropy production in the medium, we can
assign an entropy to the system itself even in nonequilibrium~\cite{seif05a}
by defining
\begin{equation}
  \label{eq:s}
  s(\tau)\equiv-\kB\ln \ps(x(\tau)).
\end{equation}
Here, the measured or calculated stationary distribution $\ps(x)$ of the
position in the steady state is evaluated along the specific trajectory
$x(\tau)$. Then the total entropy production $\st=\sm+\Delta s$ fulfills the
fluctuation theorem
\begin{equation}
  \label{eq:ft}
  \frac{P(-\st)}{P(+\st)} = e^{-\st/\kB}
\end{equation}
for any trajectory length $t$, where $P(-\st)$ is the probability of entropy
annihilating trajectories which is compared to those generating the same
positive amount of entropy.

The physical source of entropy production in our setup is the nonconservative
force $f$ which breaks detailed balance and leads to a permanent dissipation
of heat into the surrounding heat bath. Breaking of detailed balance is
quantified by the mean local velocity $\vloc(x)$, i.e., the velocity averaged
over the subset of trajectories passing $x$. With the stationary current $\js$
and probability $\ps(x)$, the mean local velocity can be expressed as
$\vloc(x)=\js/\ps(x)$. Introducing an effective potential
$\phi(x)\equiv-\ln\ps(x)$, the total force
\begin{equation}
  \label{eq:force}
  \mu_0F(x) = \vloc(x) - D_0\pd{\phi(x)}{x}
\end{equation}
splits into the local mean velocity and the gradient of the effective
potential~\cite{spec06}. From the definition of the entropy~\eqref{eq:s} it is
clear that the change of the effective potential along a stochastic trajectory
equals the change in system entropy
\begin{equation}
  \label{eq:sdelta}
  \Delta s = \kB\Delta\phi \equiv \kB[\phi(x_t)-\phi(x_0)],
\end{equation}
where $x_0$ and $x_t$ are the initial and final position of the particle,
respectively.

The change in medium entropy depends only on the initial and final position of
the particle, since in our case, due to stationarity, Eq.~\eqref{eq:sm}
simplifies to
\begin{equation}
  \label{eq:smdelta}
  T\sm(x_0,x_t) = -\Delta V + f\Delta x.
\end{equation}
Beside the sum of system and medium entropy, we can obtain an independent
expression for the total entropy production by inserting the
force~\eqref{eq:force} into Eq.~\eqref{eq:sm}. We then get after one
integration by parts and cancellation of the boundary term
\begin{equation}
  \label{eq:stot}
  \st[x(\tau)] = \frac{\kB}{D_0}\IInt{\tau}{0}{t}\dot x(\tau)\vloc(x(\tau)),
\end{equation}
which depends on the whole trajectory $x(\tau)$.


\begin{figure}[t]
  \onefigure[width=\linewidth]{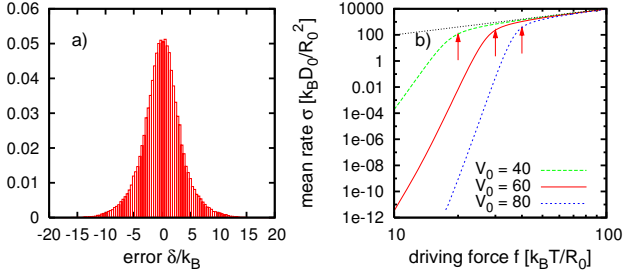}
  \caption{a) The difference $\delta\equiv\st^{(N)}-\sm-\Delta s$ after
    $t=5\unit{s}$ between the total entropy production~\eqref{eq:stotdelta}
    determined from a discretely measured trajectory and the sum of
    medium~\eqref{eq:smdelta} and system entropy~\eqref{eq:sdelta} for
    parameters $V_0=80.4\kT$ and $f=16.4\funit$. b) Mean entropy production
    rate $\sigma$ determined numerically for a potential $V(x)=(V_0/2)\cos x$
    versus driving force $f$. The dotted line indicates the limiting behavior
    $\sigma\approx f^2$ for large forces and the arrows mark the critical
    forces $\fc=V_0/(2R_0)$.}
  \label{fig:gauss}
\end{figure}

We generate a nonequilibrium steady state experimentally by driving a single
silica bead with diameter $1.85\unit{\mu m}$ along a toroidal trap with radius
$R_0\simeq2.2\unit{\mu m}$ implying a periodicity of $L=2\pi R_0$. The trap is
created by tightly focused optical tweezers rotating with a frequency of
$\trap\simeq510\unit{Hz}$~\cite{lutz06}. At this frequency, the particle
experiences a periodic driving force whenever it is ``kicked'' by the passing
laser beam but the particle cannot follow the beam directly. Since video
microscopy is not able to resolve these single kicking events [due to its
spatial ($50\unit{nm}$) and temporal ($50\unit{ms}$) resolution], the particle
is effectively subjected to a constant force $f$, thus driving it into a
nonequilibrium steady state~\cite{fauc95}. An additional periodic potential
$V(x)$ (see Fig.~\ref{fig:dist}) with depth $V_0$ is created by modulating the
intensity of the optical tweezers with an electro-optical modulator (EOM)
whose input is synchronized with the tweezers' scanning motion. The potential
$V(x)$ and the driving force
\begin{equation}
  \label{eq:f}
  f = \frac{1}{\mu_0L}\IInt{x}{0}{L}\vloc(x)
\end{equation}
are reconstructed from the measured distribution $\ps(x)$ and current
$\js$~\cite{blic07a}.

The position of the particle in polar coordinates is sampled with frequency
$20\unit{Hz}$. The deviation of the radial component $\delta
r\simeq0.06\unit{\mu m}$ with $\delta r/R_0<3\%$ is small enough to justify
the assumption of an effectively one-dimensional motion. We record one long
trajectory from which we determine the stationary probability $\ps(\al)$ of
the angular position $\al$ shown in Fig.~\ref{fig:dist}. The trajectory is
then divided in overlapping segments of $N$ points such that the angles
$\al_i$ with $1\leqslant i\leqslant N$ form discrete trajectories. The total
entropy production along one discrete trajectory is calculated from
Eq.~\eqref{eq:stot} as
\begin{equation}
  \label{eq:stotdelta}
  \st^{(N)} = \kB\frac{\js R_0^2}{D_0}
  \sum_{i=2}^{N-1}\frac{\al_{i+1}-\al_{i-1}}{2\ps(\al_i)}.
\end{equation}
In Fig.~\ref{fig:gauss}a, the deviation $\delta\equiv\st^{(N)}-\sm-\Delta s$
of the total entropy production~\eqref{eq:stotdelta} from the independently
measured medium entropy production~\eqref{eq:smdelta} and the entropy
production of the system itself~\eqref{eq:sdelta} is shown. This deviation is
a Gaussian centered around zero with a standard deviation of $3.5\kB$, which
corresponds to a relative error $<3\%$ given the mean
$\mean{\st}\simeq121.3\kB$. This small error shows that the discretization of
the stochastic velocity $\dot x$ within the integral Eq.~\eqref{eq:stot} is a
very good approximation even for a time resolution of $50\unit{ms}$. This is
not obvious {\it a priori} due to the mathematically nondifferentiable
stochastic paths $x(\tau)$.

\begin{figure}[t]
  \onefigure[width=\linewidth]{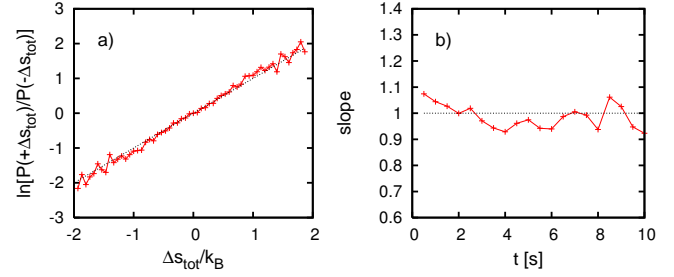}
  \caption{a) Test of the fluctuation theorem~\eqref{eq:ft} for
    $t=0.75\unit{s}$, where the logarithm $\ln[P(+\st)/P(-\st)]$ is plotted
    versus $\st$. The dashed line indicates the expected slope of one and in
    b) the effective slope is plotted with increasing trajectory lengths.
    (Parameters: $V_0=65.6\kT$, $f=14.2\funit$)}
  \label{fig:ft}
\end{figure}

In Fig.~\ref{fig:ft}, the data are plotted in the form of the fluctuation
theorem~\eqref{eq:ft} for different trajectory lengths $t$. The fluctuation
theorem can only be tested directly in a small window of the histogram
centered around zero due to the need for negative events. Since the
distribution $P(\st)$ shifts towards larger values with increasing time,
negative values of $\st$ become less probable and the statistics is good
enough only for relatively small times.

The driving force $f$ tilts the potential $V(x)$ and lowers the potential
barrier which at the critical force $\fc$ vanishes and deterministic running
solutions start to exist. For a cosine potential, the critical force turns out
to be $\fc=V_0/(2R_0)$. In Fig.~\ref{fig:gauss}b, the mean entropy production
rate
\begin{equation}
  \label{eq:rate}
  \sigma \equiv \mean{\partial_t\st} = \frac{\kB}{D_0}\mean{\vloc^2}
  = \frac{\kB}{D_0}\js^2R_0^2\IInt{\al}{0}{2\pi}\ps^{-1}(\al)
\end{equation}
is plotted versus the driving force for three potential depths $V_0$. The two
limiting cases
\begin{equation}
  \sigma \approx
  \begin{cases}
    (\mu_0/T) f^2   & (f\gg\fc) \\
    (\Delta w/T)\rk & (f\ll\fc)
  \end{cases}
\end{equation}
are understood easily. For large forces $f\gg\fc$, the potential becomes
irrelevant and the particle diffuses freely with drift velocity $\propto f$.
The mean entropy production rate in this case becomes $(\mu_0/T)f^2$. For
small forces $f\ll\fc$, the particle is mostly trapped within one minimum from
which it escapes with the Kramers' rate $\rk=r_0\exp\{(-V_0+fL/2)/\kT\}$ where
$r_0$ is the attempt rate~\cite{hang90}. If the particle moves to the next
minimum, the driving force has spent the work $\Delta w=fL$.

\begin{figure}[t]
  \onefigure[width=\linewidth]{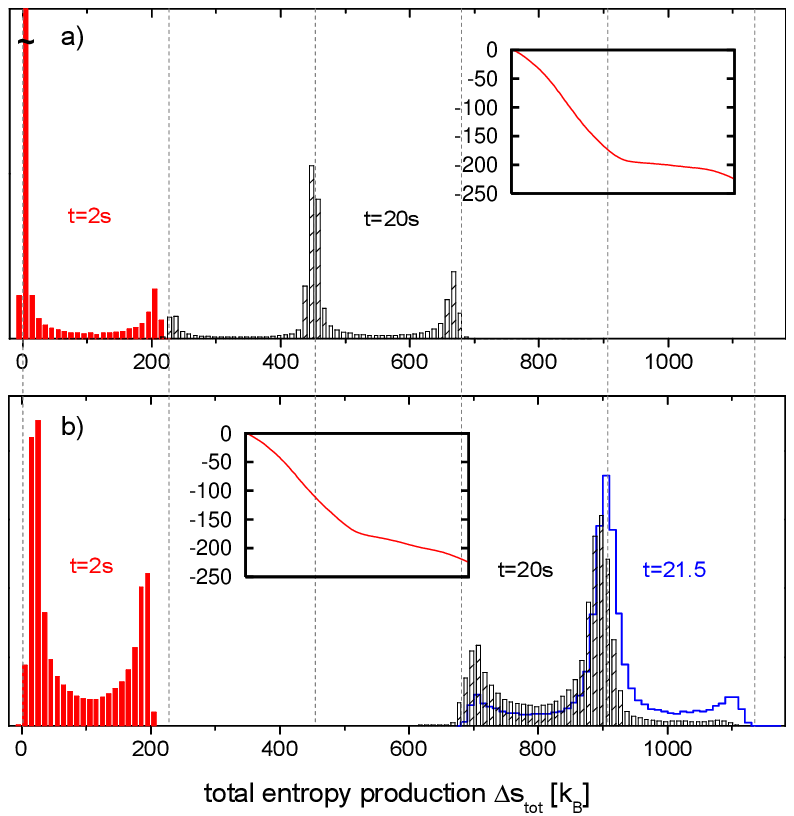}
  \caption{Histograms of total entropy production $\st$ at the crossover from
    locked to running state: a) $V_0=80.4\kT$ and $f=16.4\funit$ which
    corresponds to the critical force, b) $V_0=61.7\kT$ and the same force
    $f=16.4\funit$ corresponding to the running state. In both graphs, the
    left histograms (closed bars) are recorded for trajectories of length
    $t=2\unit{s}$, the right histograms (hatched bars) for $t=20\unit{s}$. In
    addition, b) shows as a solid line for $t=21.5\unit{s}$ an almost
    symmetric histogram (with respect to its peak). The dashed vertical lines
    are separated by the amount of work $\Delta w=fL$ spent by the driving
    force $f$ in one revolution. The insets display the tilted potentials
    $V(x)-fx$ for the two cases.}
  \label{fig:hist}
\end{figure}

This crossover can also be observed in the histograms of $\st$. In
Fig.~\ref{fig:hist}a, the case with the critical force $f\simeq\fc$ is
plotted. The work spent by the driving force is the product of force times the
displacement of the particle such that trajectories corresponding to $n$
revolutions of the particle lead to an entropy production peaked at the dashed
vertical lines with $\st=n\Delta w/T$. In Fig.~\ref{fig:hist}b, the potential
depth $V_0$ has been decreased at the same driving force, leading to $f>\fc$.
In this case the distribution of $\st$ starts to ``run'', i.e., the peak
positions are not fixed anymore at the vertical lines. However, this is where
they still reach their maximum as demonstrated by the rightmost histogram for
$t=21.5\unit{s}$.


\begin{figure}[t]
  \onefigure[width=\linewidth]{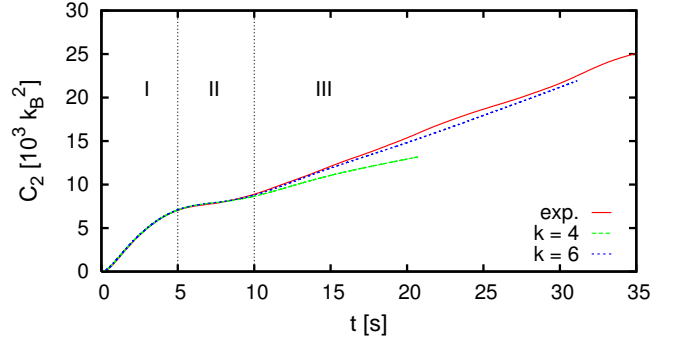}
  \caption{Comparison of a) the variance $C_2(t)$ and b) the skewness $C_3(t)$
    between experimental data and the solution of Eq.~\eqref{eq:mom}. The
    experimental data has been obtained from 120.000 trajectories for
    $V_0=65.6\kT$ and $f=14.2\funit$. $k$ is the number of Fourier
    coefficients used for parameterizing the actual potential $V(x)$, see the
    inset of Fig.~\ref{fig:dist}. The three regimes I, II, and III are
    discussed in the main text.}
  \label{fig:mom}
\end{figure}

The probability distribution $P(\st)$ of the total entropy production is a
nonuniversal function distinctly non-Gaussian and evolving in time. In order
to compare the experimental data to theory, we need to calculate $P(\st)$
independently. In Ref.~\cite{spec05a} an equation governing the time evolution
of the joint probability distribution $\rho(x,r=\st,t)$ was
derived~\footnote{Eq.~(22) in Ref.~\cite{spec05a} was derived originally for
  the so called ``housekeeping heat'' $Q_\mathrm{hk}$.  Since in our case
  $\st=Q_\mathrm{hk}/T$, this equation holds also for the total entropy
  production.}. The conditional moments $m_n(x,t)\equiv\Int{r}r^n\rho(x,r,t)$
are the contribution of all trajectories ending in $x$ at time $t$ to the
moments
\begin{equation}
  \label{eq:raw}
  M_n(t) \equiv \mean{(\st)^n} = \IInt{x}{0}{L}m_n(x,t)
\end{equation}
of $P(\st)$. Following Ref.~\cite{spec05a}, the time evolution of the
conditional moments reads
\begin{equation}
  \label{eq:mom}
  \pd{m_n(x,t)}{t} = \hat L m_n(x,t) + S_n(x,t)
\end{equation}
where $\hat L\equiv-\partial_x\left[\mu_0F(x) - D_0\partial_x\right]$ is the
Fokker-Planck operator determining the Brownian motion of the particle. The
source term
\begin{equation*}
  S_n = -n\kB\left[2\pd{}{x}\vloc - \frac{\vloc^2}{D_0}\right] m_{n-1} 
  + n(n-1)\kB^2\frac{\vloc^2}{D_0}m_{n-2}
\end{equation*}
couples the evolution of the $n$th conditional moment $m_n$ to conditional
moments of lower order where $m_0(x)=\ps(x)$ is the stationary solution of
$\hat Lm_0=0$.

For the numerical calculation, we use the experimentally obtained potential
$V(x)$, driving force $f$, stationary current $\js$, and probability
distribution $\ps(x)$ for the run shown in Fig.~\ref{fig:dist} measured in the
vicinity of the critical force $f\simeq\fc$. Because of the periodicity of
$m_0(x+L)=m_0(x)$, $m_n(x)$ as well as the source terms $S_n(x)$ are also
periodic for $n\geqslant0$. Eq.~\eqref{eq:mom} is therefore easily solved
numerically in Fourier space which we have done for the first two moments
$M_{1,2}$. From these (i) the mean $M_1$ and (ii) the variance $C_2\equiv
M_2-M_1^2$ are obtained. The mean $M_1(t)=\sigma t$ is a straight line from
which we can extract $\sigma$.  First, we fit the experimental data with
$\sigma\simeq17.15\kB/\mathrm{s}$.  Second, we fit the numerical
solution~\eqref{eq:raw} leading to $\sigma\simeq17.12\kB/\mathrm{s}$ in
excellent agreement with the experimental data. The latter value of $\sigma$
is also obtained from Eq.~\eqref{eq:rate} which only involves the measured
distribution $\ps(x)$ and current $\js$.

In Fig.~\ref{fig:mom}, we plot the variance $C_2$ together with its
experimental counterpart. The time-dependence of the variance $C_2(t)$
resembles roughly that of the mean square displacement. During the first
regime (I) in Fig.~\ref{fig:mom}a, the particle explores its vicinity until it
reaches on average the potential barrier. While surmounting the barrier (II),
the spreading of the distribution slows down and then again increases
approximately linearly (III). Analyzing the data at the critical force
$f\simeq\fc$ demonstrates the sensitivity of the entropy production. The slope
in regime III is especially sensitive with respect to both the potential and
the driving force. Therefore the force $f$ used in the numerical calculations
has been fitted with value $f\simeq13.9\funit$ to match the experimentally
determined $C_2(t)$. This corresponds to a deviation of about $2\%$ compared
to the value $f\simeq14.2\funit$ calculated from Eq.~\eqref{eq:f}, which is
well within the estimated error of the force. The accuracy of the potential's
parameterization is controlled by the number $k$ of Fourier coefficients used
leading to a better agreement with higher value for $k$. Despite the good
agreement for the mean and variance, the numerical calculation of higher
cumulants shows an increasing sensitivity with respect to the accuracy of the
measured quantities needed as input.


In summary, we have measured experimentally the distribution of the total
entropy production caused by driving a colloidal particle in a toroidal
geometry. The system exhibits a transition from exponentially small to
quadratic mean entropy production rate depending on the ratio $fL/V_0$ between
driving force $f$ and potential depth $V_0$, which can be seen in the
histograms of the entropy production as well. The time evolution of the
moments of the total entropy production is described by a differential
equation. The procedure outlined above becomes less reliable for higher
cumulants due to accumulating errors with increasing trajectory length. For
long trajectories, direct calculation of the asymptotic large deviation
function of the entropy production rate seems preferable. Its extraction from
experimental data, however, might pose a challenge.



\end{document}